\documentclass[prl,aps,reprint,floatfix,superscriptaddress]{revtex4-2}

\usepackage{amsmath}
\usepackage{amssymb}
\usepackage{amsfonts}
\usepackage{float}
\usepackage[colorlinks=true,citecolor=blue,linkcolor=magenta]{hyperref}
\usepackage{graphicx}
\usepackage{bm}
\usepackage{bbold}
\usepackage{color}
\usepackage[dvipsnames]{xcolor}
\usepackage{silence}
\DeclareMathAlphabet\mathbfcal{OMS}{cmsy}{b}{n}
\newcommand{\avg}[1]{{\langle#1\rangle}}

\newcommand{\tr}{{\mathrm{tr}}}
\newcommand{\im}{{i}}
\newcommand{\dg} {{\dagger}}
\newcommand{\pd} {{\phantom\dagger}}
\newcommand{\ai}[1] {{a_{#1}^{\pd}}}
\newcommand{\aid}[1] {a_{#1}^\dg}

\newcommand{\ci}[1] {{c_{#1}^{\pd}}}
\newcommand{\cid}[1] {c_{#1}^\dg}
\newcommand{\hc} {\text{h.c.}}

\renewcommand{\Im}{\operatorname{Im}}
\newcommand{\ql}{\mathcal{L}}

\newcommand{\qs}{\mathcal{S}}
\newcommand{\qr}{\mathcal{R}}

\newcommand{\qd}{\mathcal{D}}
\newcommand{\qf}{\mathcal{F}}

\newcommand{\NS}{N_\qs}

\newcommand{\Iav}{I_\diamondsuit}

\newcommand{\U}{U}
\newcommand{\vS}{v_\qs}
\newcommand{\VS}{V_\qs}

\newcommand{\oxox}{{\large$\circ$$\bullet$$\circ$$\bullet$}}
\newcommand{\xoox}{{\large$\bullet$$\circ$$\circ$$\bullet$}}
\newcommand{\xxoo}{{\large$\bullet$$\bullet$$\circ$$\circ$} }
\newcommand{\xoxo}{{\large$\bullet$$\circ$$\bullet$$\circ$}}

\newcommand{\xxox}{{\large$\bullet$$\bullet$$\circ$$\bullet$}}
\newcommand{\xxxo}{{\large$\bullet$$\bullet$$\bullet$$\circ$}}
\newcommand{\oxxx}{{\large$\circ$$\bullet$$\bullet$$\bullet$}}

\newcommand{\NESS}{\bar{\rho}_\qs}
\newcommand{\HR}{\breve{H}_\qs}

\renewcommand{\[}{\begin{equation}}
\renewcommand{\]}{\end{equation}}
\usepackage{soul}

\begin{document}
\title{The confluence of fractured resonances at points of dynamical, many--body flare}
\author{Bitan De}\affiliation{Institute of Theoretical Physics, Jagiellonian University, Łojasiewicza 11, 30-348 Kraków, Poland}
\author{Gabriela W\'ojtowicz}
\affiliation{Institute of Theoretical Physics, Jagiellonian University, Łojasiewicza 11, 30-348 Kraków, Poland}
\affiliation{Doctoral School of Exact and Natural Sciences, Jagiellonian University, Łojasiewicza 11, 30-348 Kraków, Poland}
\affiliation{Biophysical and Biomedical Measurement Group, Microsystems and Nanotechnology Division, Physical Measurement Laboratory, National Institute of Standards and Technology, Gaithersburg, Maryland 20899, USA}
\author{Marek M. Rams}
\email{marek.rams@uj.edu.pl}
\affiliation{Institute of Theoretical Physics, Jagiellonian University, Łojasiewicza 11, 30-348 Kraków, Poland}
\affiliation{Mark Kac Center for Complex Systems Research, Jagiellonian University,  Łojasiewicza 11, 30-348 Kraków, Poland}
\author{Michael Zwolak}
\email{mpz@nist.gov}
\affiliation{Biophysical and Biomedical Measurement Group, Microsystems and Nanotechnology Division, Physical Measurement Laboratory, National Institute of Standards and Technology, Gaithersburg, Maryland 20899, USA}
\author{Jakub Zakrzewski}
\email{jakub.zakrzewski@uj.edu.pl}
\affiliation{Institute of Theoretical Physics, Jagiellonian University, Łojasiewicza 11, 30-348 Kraków, Poland}
\affiliation{Mark Kac Center for Complex Systems Research, Jagiellonian University, Łojasiewicza 11, 30-348 Kraków, Poland}

\begin{abstract}
Resonant transport occurs when there is a matching of frequencies across some spatial medium, increasing the efficiency of shuttling particles from one reservoir to another. 
We demonstrate that in a periodically driven, many--body titled lattice, there are sets of spatially fractured resonances. 
These ``emanate'' from two essential resonances due to scattering off internal surfaces created when the driving frequency and many--body interaction strength vary, a scattering reminiscent of lens flare. 
The confluence of these fractured resonances dramatically enhances transport.
At one confluence, the interaction strength is finite and the essential resonance arises due to the interplay of interaction with the counter--rotating terms of the periodic drive. 
We discuss the origin and structure of the fractured resonances, as well as the scaling of the conductance with system parameters. 
These results furnish a new example of the richness of open, driven, many--body systems. 
\end{abstract}

\date{\today}

\maketitle
 
Closed, one--dimensional lattices held at a tilt provide a simple realization of many--body localization in clean,  quantum systems~\cite{van_nieuwenburg_bloch_2019,schulz_stark_2019}. 
The localized dynamics were initially linked to the fragmentation of the Hilbert space and the conservation of a global dipole moment~\cite{khemani_localization_2020,sala_ergodicity_2020}. 
This approach was further refined and expanded by numerous theoretical and experimental studies, specifically in the context of cold atoms and trapped atomic ions~\cite{taylor_experimental_2020,chanda_coexistence_2020,yao_many-body_2021,yao_nonergodic_2021,doggen_stark_2021,morong_observation_2021,scherg_observing_2021,kohlert_exploring_2023}. 

While the closed system is relatively well understood, much less is known for tilted lattices in contact with an environment. 
Among other considerations, it is interesting from the context of quantum transport where much experimental and theoretical progress has been made in one--, and quasi--one--, dimensional atomic quantum systems~\cite{brantut_conduction_2012,chien_bosonic_2012,brantut_thermoelectric_2013,chien_interaction-induced_2013,kohler_driven_2005,krinner_observation_2015,krinner_mapping_2016,krinner_two-terminal_2017,hausler_scanning_2017,gruss_energy-resolved_2018,karevski_quantum_2009, prosen_exact_2011, prosen_open_2011, znidaric_spin_2011, karevski_exact_2013,bertini_finite-temperature_2021,landi_nonequilibrium_2022}.  
In quantum transport, the environments coupled to the lattice induce a particle imbalance that results in a current  across the lattice.  
Current--carrying particles explore a wide range of energy spectra of the junction, thus the current measured across the junction can be linked to its spectral properties~\cite{kiczynski_engineering_2022,wang_experimental_2022}. 
In  tilted lattices, recent studies have considered time--independent transport in non--interacting~\cite{pinho_bloch_2023} and interacting~\cite{mendoza-arenas_giant_2022} systems, as well as time--dependent transport in non--interacting lattices in contact with non--Markovian fermionic reservoirs~\cite{de_transport_2023}. 
These works demonstrate interesting phenomena, such as rectification, and highlight challenges in numerical simulation. 

\begin{figure}[t]
\includegraphics[width=0.75\columnwidth]{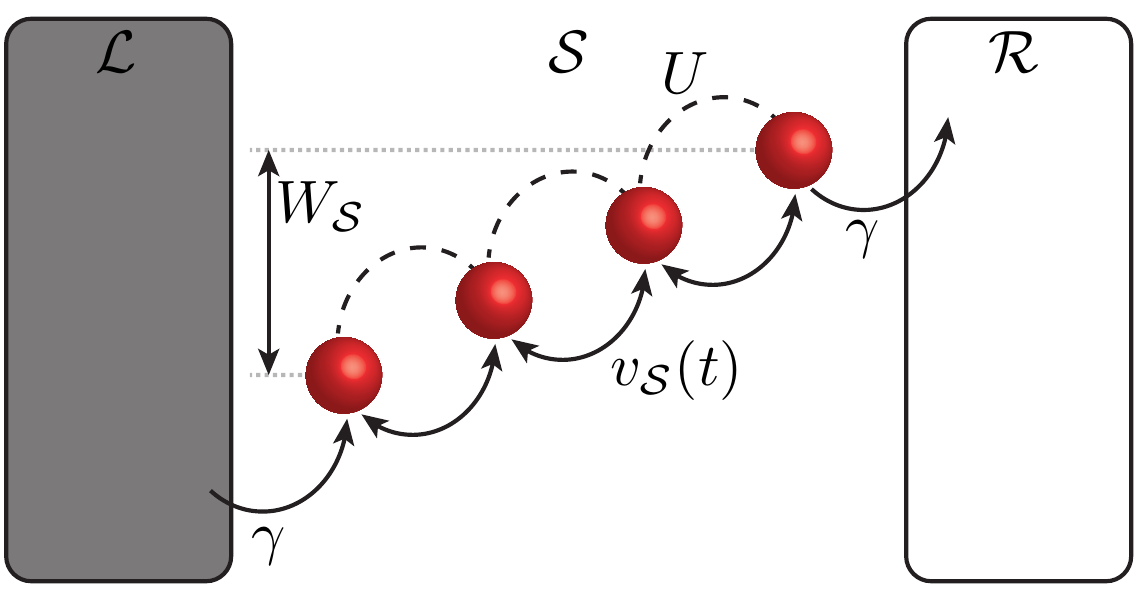}
   \caption{{\bf Many--body, driven, tilted lattice.} 
   A spinless fermion lattice $\mathcal{S}$ of $\NS$ sites is held at a total tilt $W_\qs = (\NS-1) \Delta$ and placed between Markovian reservoirs, $\mathcal{L}$ and $\mathcal{R}$, with injection and depletion, respectively, at rate $\gamma$. 
   Fermions interact with a nearest--neighbor, density--density interaction of strength $U$, and tunneling coupling $\vS (t)$ that is periodically driven. 
   The triple of elements---the periodic modulation, the interactions, and the tilt---gives rise to the confluence of resonance and a bulk ballistic channel.}
\label{fig:schematic}
\end{figure}

In this Letter, we consider transport in periodically driven, interacting tilted lattices, as in  Fig.~\ref{fig:schematic}. 
We show that the combination of modulation, many--body interaction, and tilt leads to sets of fractured resonances. 
These resonances come together in a pair of confluence points where the conductance markedly increases. 
We will discuss the fractured pathways, the current scaling in different regimes, and the emergence of a bulk channel. 

The one--dimensional lattice consists of $\NS$ sites where Markovian injection and depletion on the first and last site, respectively, induce transport through the system~$\qs$. 
This setup is essentially the large bias and bandwidth limit of $\qs$ between left ($\ql$) and right ($\qr$) reservoirs, as depicted in Fig.~\ref{fig:schematic}.
The time evolution of the density matrix, $\rho_\qs$, is governed by the equation of motion, 
\[ \label{eq:eom}
    \dot{\rho}_\qs = -\im \left[ H_\qs(t), \rho_\qs \right] + \qd_\qs \left[ \rho_\qs \right] \equiv \ql_\qs (t) \left[ \rho_\qs \right],
\]
where $H_\qs (t)$ is the time--dependent system Hamiltonian and $\qd_\qs$ provides the Markovian injection and depletion on the boundaries. 
These two contributions combine to give the time--dependent Lindbladian $\ql_\qs (t)$. 

In terms of the fermionic creation (annihilation) operators, $\cid{j}$ ($\ci{j}$), on site $j$, as well as the number operator, $n_j=\cid{j} \ci{j}$, the system Hamiltonian is 
\begin{eqnarray} \label{eq:HS}
    H_\qs(t) &=&  \sum_{j=1}^{\NS} j \Delta n_j +  \sum_{j=1}^{\NS-1} \U n_j n_{j+1} \nonumber \\ 
    && + \vS (t) \sum_{j=1}^{\NS-1}  \left(\cid{j} \ci{j+1} + \hc\right), 
\end{eqnarray}
where $\vS (t) = \VS \cos ( \omega t )$ and $\Delta$ is the site--to--site tilt. 
The periodically--driven, nearest--neighbor coupling has amplitude $\VS$ and frequency $\omega$, i.e., a driving period $T=2\pi/\omega$, while $U$ is the density--density interaction strength between neighboring sites. 
The boundary terms are given by the Markovian contribution,  
\begin{eqnarray} \label{eq:boundary}
    \qd_\qs [ \rho_\qs ] & = & \gamma \Bigl( \cid{1} \rho_\qs \ci{1} - \frac{1}{2} \left\{ \ci{1} \cid{1} , \rho_\qs \right\}  \Bigl) \nonumber  \\
    && + \gamma \Bigl( \ci{\NS} \rho_\qs \cid{\NS} - \frac{1}{2} \left\{ \cid{\NS} \ci{\NS} , \rho_\qs \right\} \Bigl), 
\end{eqnarray}
that gives injection at the first site and depletion at the last site of $\qs$. 
The driving in Eq.~\eqref{eq:boundary} can be relaxed to having both injection and depletion at the boundaries~\cite{landi_nonequilibrium_2022} but the physical interpretation of such Markovian driving is elusive. 
Since, the confluence of resonances is a bulk effect, changing the driving in Eq.~\eqref{eq:boundary} plays a minor role. 
We will focus on the current supplied by Eq.~\eqref{eq:boundary}, as well as the resonance structure for a fixed tilt $\Delta$ at small $\VS$ and $\gamma$. 
This includes $\VS / \Delta \ll 1$ to minimally smear bare atomic levels. 

For the periodic model in Eq.~\eqref{eq:eom}, there exist a periodic steady state solution, i.e.,  $\rho_\qs(t+T)=\rho_\qs(t)$. 
The steady state can be found directly by identifying non--decaying eigenvector of the Floquet map, 
\[
\qf_\qs = \mathcal{T} e^{\int_0^T \mathcal{L_\qs}(t)dt} ,
\label{floquetmap}
\]
where $\mathcal{T}$ is the time--ordering operator and $\NESS = \qf_\qs \left[ \NESS \right]$ is the $T$--periodic stationary state. 
Alternatively, the Floquet approach provides an elegant method of treating the periodic drive~\cite{shirley_solution_1965} by looking for a solution in the Fourier expansion neglecting quickly oscillating terms if possible.

\begin{figure}
\includegraphics[width=\columnwidth]{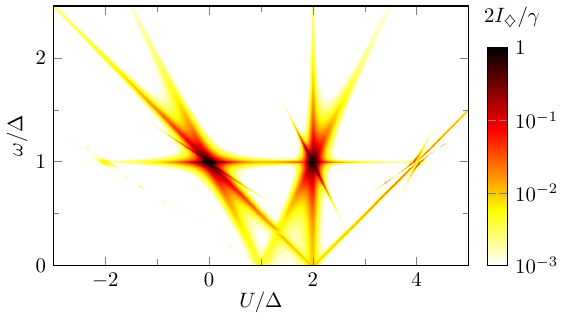}
   \caption{{\bf Confluence of resonance.} 
   The current (color scale, right) versus the driving frequency $\omega$ and many--body interaction strength $U$ indicate the location of resonances in the lattice of $\NS=4$ sites. 
   These resonances come together at two confluence points at interaction $U=0$ and $U=2 \Delta$, both for driving $\omega=\Delta$.
   At $U=2 \Delta$, the resonant line $\omega=\Delta$ coincides with the driving--independent $U=2 \Delta$ resonance and the resonances at $\omega=U-\Delta$ and $\omega=3\Delta - U$. 
   At $U=0$, the $\omega=\Delta$ resonant line merge with strong $\omega=\Delta-U$ and $\omega=\Delta-U/2$ resonances, and a very fine contribution at $\Delta-U/3$. 
   The data are computed at $\gamma=0.01 \Delta$ and $\VS=0.1 \Delta$. }
\label{fig:confluence}
\end{figure}

\begin{figure*}[t!]
\includegraphics[width=0.95\textwidth]{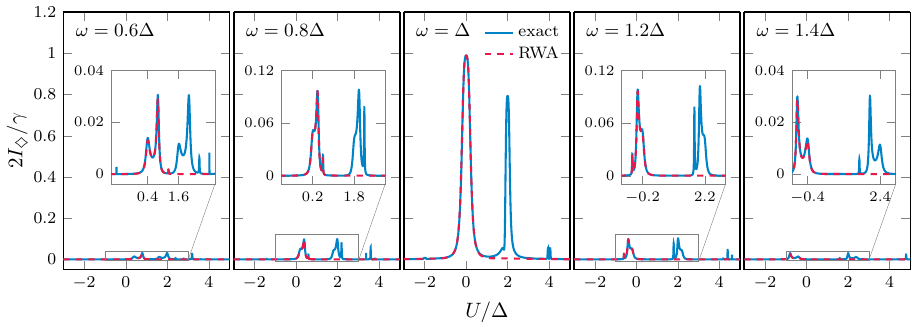}
\caption{{\bf Resonance structure.} 
    The mean current over a cycle, Eq.~\eqref{eq:MeanCurrent}, versus $U/\Delta$ for different drives $\omega$ through a lattice of $\NS=4$ sites. 
    The blue solid lines show the exact mean current, which exhibits a rich resonance structure. 
    At the drive $\omega=\Delta$, the separate resonances merge at $U=0$ and $U=2 \Delta$. 
    This confluence leads to a large enhancement of the current. 
    The red dashed lines show the rotating wave approximation (RWA), which accurately captures the small $U$ resonances but not those around $U=2 \Delta$. 
    The setup is the same as Fig.~\ref{fig:confluence}. 
}
\label{fig:resonantstructure}
\end{figure*}

In this paper, we focus on transport properties, quantified by the mean current over the driving period, 
\[
\begin{split} \label{eq:MeanCurrent}
    \Iav=\frac{1}{T}\int_{0}^{T} I_j(t+\tau) d\tau .
\end{split}
\]
The steady--state is independent of the time $t$ and, due to particle conservation, independent of the interface $j$. 
The current for a specific time and interface is  
\[ \label{curi}
    I_j (t) =2 \vS (t) \Im \avg{\cid{j} \ci{j+1}}_t ,
\]
where $\avg{A}_t=\tr[A\NESS(t)]$ indicates the steady--state expectation value.

The driven interacting model, Eq.~\eqref{eq:eom}, exhibits a rich structure of resonances. 
In Fig.~\ref{fig:confluence}, the resonances are visible as tracks of enhanced current when $U$ and $\omega$ vary. 
Multiple resonances meet at two high symmetry points: at $U=0$ (a non--interacting system) and $U=2 \Delta$, both for $\omega=\Delta$.
These two points---essential resonances or confluence points---form due to the merger of several resonances, as seen in Fig.~\ref{fig:confluence}. 
The confluence effect is not additive for the current, showing non--trivial interference of the resonance pathways, as detailed in Fig.~\ref{fig:resonantstructure}.  

To explain the resonances, we examine the Hamiltonian, Eq.~\eqref{eq:HS}, in the rotating frame ($\ai{j} \equiv e^{\im j \omega  t} \ci{j}$), 
\begin{eqnarray} \label{eq:HR}
    \HR (t) &=&  \sum_{j=1}^{\NS} j\left(\Delta-\omega\right) n_j + \sum_{j=1}^{\NS-1} \U n_j n_{j+1} \nonumber \\ 
    && +\frac{\VS}{2} \sum_{j=1}^{\NS-1} \left[\aid{j} \ai{j+1} (1+e^{-2 \im \omega t}) + \hc \right] \\
    &\equiv& H_0 + H_1 e^{-2\im \omega t} + H_1^\dagger e^{2\im\omega t},  \nonumber
\end{eqnarray}
where $H_1=\sum_{j} (\VS/2) \aid{j} \ai{j+1}$ is a quickly rotating term that is dropped in the RWA. 
If we associate the driving frequency $\omega$ with photon processes, the model in Eq.~\eqref{eq:HR} has a single--photon resonance at $\omega=\Delta$. 
Figure~\ref{fig:resonantstructure} shows the comparison between the RWA and exact current. 
The RWA is a good approximation for weak coupling $U$ but it fails completely at the confluence point $U=2 \Delta$. 

\paragraph{Confluence of resonance at $U=0$.} 
Since the RWA captures this confluence, we can  
safely drop the counter--rotating terms in Eq.~\eqref{eq:HR} making the lattice a uniform, non--interacting lattice between two Markovian reservoirs when $\omega=\Delta$. 
The current is thus~\cite{karevski_quantum_2009,znidaric_matrix_2010}
\[ \label{eq:current_RWA}
\Iav = \frac{\gamma}{2} \frac{\VS^2}{\VS^2 + \gamma^2 },
\]
which uses the RWA hopping of $\VS/2$. 
When $\gamma \ll \VS$, the current approaches  $\gamma/2$.
In this limit, the lattice itself has effectively zero resistance as the uniform, non--interacting lattice forms a ballistic channel.
The setup is thus well described by two contributions in series, one at each of the  interfaces~\cite{gruss_landauers_2016}. 
When $\VS \ll \gamma$, the current is $\VS^2/2\gamma$. This dependence occurs due to the overdamping of the first and last sites in direct analogy to Kramers' turnover within extended reservoir approaches for quantum~\cite{gruss_landauers_2016,elenewski_communication_2017,gruss_communication_2017,zwolak_analytic_2020,zwolak_comment_2020,wojtowicz_dual_2021,elenewski_performance_2021,wojtowicz_accumulative_2023} and classical~\cite{velizhanin_driving_2011,chien_tunable_2013,chien_thermal_2017,chien_topological_2018} systems. 

Around confluence, the resonances for small $U$ are also well--reproduced by the RWA, see Fig.~\ref{fig:resonantstructure}, including for $\omega$ away from $\Delta$. However, {\it these are fractured resonances}, as the lattice no longer has a uniform  frequency facilitating transport. 
To understand the fracturing, we analyze the state structure of the time--independent RWA Hamiltonian, $H_0$. 
As in Fig.~\ref{fig:resonantstructure}, we focus on $\NS=4$.  
One of the fractured resonances, at $\omega=\Delta$, is independent of the many--body interaction.
It has a homogeneous, single--particle channel, similar to $U=0$ (a closely related model---XXZ---remains ballistic for $|U| < \VS$~\cite{prosen_exact_2011}).
The other channels, though, occur at $\omega=\Delta-U$ and $\omega=\Delta-U/2$, and a very fine resonance at $\omega=\Delta-U/3$.

Considering $\omega=\Delta-U$, there are single photon transitions between Fock states \xxoo and \xoxo{ }(using {\large $\bullet$} for occupied and  {\large $\circ$} for unoccupied sites). 
These states have atomic frequencies $3 (\Delta-\omega) +U$ and $4 (\Delta-\omega)$ in the rotating frame, respectively, which become resonant when $\omega=\Delta-U$. 
The resonance is fractured as for the transport to occur across the whole system, fermions must overcome a barrier to go next into \xoox, which then allows the particle into the right reservoir and to contribute to the NESS current. 
The three particle sector, via the states \xxxo{ }and \xxox, also contributes to the fractured resonance via a single--photon process. 

The other two fractured resonances require two-- and three--photon processes, respectively. 
For $\omega=\Delta-U/2$, the transition is between \xxoo and \xoox, and has a magnitude proportional to $\VS^2$, which is suppressed compared to the first--order resonant coupling.
For $\omega=\Delta-U/3$, the transition is between \xxoo and \oxox, which is further suppressed with a higher power of $\VS$. 

\paragraph{Confluence of resonance at $U=2 \Delta$.} 
When the interaction $U$ becomes strong, the assumptions for the RWA in Eq.~\eqref{eq:HR} no longer hold (for the review on the cases where RWA breaks down, see Ref.~\cite{frisk_kockum_ultrastrong_2019}). 
The resonance originates from the the counter--rotating terms. 
To approximate the solution, we move to the rotating frame to eliminate the tilt and the density--density interaction. 
Since these terms commute, we can rotate out these terms independently arriving at Eq.~\eqref{eq:HR} and then applying the unitary transformation ${\cal U}=\exp \{2i\omega t \sum_{j=1}^{\NS-1} n_j n_{j+1}\}$, where we choose $2\omega$ frequency to target rotations around $U=2\Delta=2\omega$. 
After the transformation and dropping  fast rotating terms we obtain
\begin{eqnarray} \label{eq:Heff}
H_{\rm eff} &=&  \sum_{j=1}^{\NS} j\left(\Delta-\omega\right) n_j + \sum_{j=1}^{\NS-1} (\U-2\omega) n_j n_{j+1}  \\ 
    &+& \frac{\VS}{2} \sum_{j=1}^{\NS-1} \left[\aid{j} \ai{j+1} [1-(1-n_{j-1})n_{j+2}] + \hc \right],\nonumber  
\end{eqnarray}
with $n_0=0$ and $n_{\NS+1}=0$. 
Note that this effective Hamiltonian has a very similar structure to $\HR$ in Eq.~\eqref{eq:HR} with strongly reduced interactions but with density--dependent tunnelings resembling the situation in Hubbard models 
\cite{Hirsch94,Dutta15}. The similarity between Eq.~\eqref{eq:HR} and Eq.~\eqref{eq:Heff} suggests a comparable fracturing of resonances around $U=0$ and $U=2\Delta$. 

To understand the resonances, let us examine the energy degeneracy in Eq.~\eqref{eq:Heff}. 
For instance, the counter--rotation connects \xxoo and \xoxo, where the latter can then resonantly ($\omega=U-\Delta$ resonance) shuttle one particle to the right boundary, creating a full path through the lattice.
The connectivity, where neighboring particles with a many--body interaction  can separate, allows the lattice to stay open and efficiently transport particles across the system. 
At one of the other fractured resonances, $U=2 \Delta$, a two--photon resonance between \xxoo and \xoox{ }facilitates transfer across the lattice for arbitrary $\omega$. The linking of all the resonances: $\omega=U-\Delta$, $\omega=3\Delta - U$, $U=2 \Delta$, and $\omega=\Delta$, at confluence leads to a giant increase of the current.

\paragraph{Rectification.} 
While we observe a confluence at $U=2 \Delta$, no such effect occurs for $U=-2\Delta$, although the state structure seems similar. 
This is because $H_1$ in Eq.~\eqref{eq:HR} facilitates transport to higher frequency states to the left, while $H_1^\dagger$ facilitates transport to lower frequencies to the right. 
This breaks the spatial symmetry of the lattice, where counter--rotation permits $U=2\Delta$ to carry a current from left to right, matching the Markovian injection and depletion in Eq.~\eqref{eq:eom}.
For $U=-2\Delta$, counter--rotation permits a current from right to left but this is opposite to the induced current in Eq.~\eqref{eq:eom}, and the system fills into a blocked state \xxxo. 
Swapping the Markovian injection and depletion, while holding the tilt in the positive direction, results in the confluence at $U=-2\Delta$ and the blocked state \oxxx{ }at $U=2 \Delta$; with further tunneling prevented by the density--dependent projection in the second line of Eq.~\eqref{eq:Heff}.
The system thus displays strong rectification, although its origin is different from the rectification in {\it time--independent} lattices~\cite{landi_flux_2014, balachandran_perfect_2018,mendoza-arenas_giant_2022}.
 
\begin{figure}
\includegraphics[width=\columnwidth]{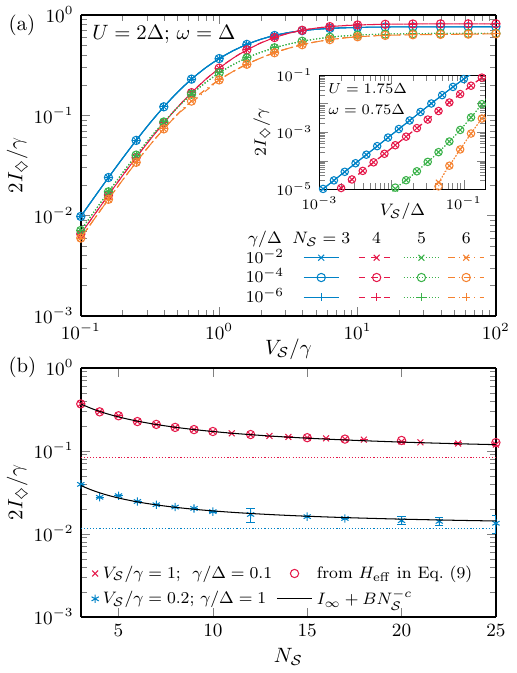}
\caption{{\bf Current at the many--body confluence $U=2\Delta$}. 
(a) The current versus $V_\qs/\gamma$ reveals the scaling limit for small $\VS$ where the current is proportional to $\VS^2 / \gamma$, similar to $U=0$ case, Eq.~\eqref{eq:current_RWA}.  
The inset shows qualitatively different behavior away from the confluence, where the current vanishes with a higher power of $\VS$ and has a strong dependence on $\NS$.
(b) The current versus $\NS$ for two sets of parameters. The fits to $\Iav= I_\infty + B \NS^{-c}$ are shown as solid, black lines. The asymptotic values, $I_\infty$ (horizontal, dotted lines), indicate a significant bulk resonant channel. When $\VS$ and $\gamma$ are small compared to $\Delta$ (upper curve), the full numerical simulations (red crosses) agree very well with those from the time-independent $H_{\rm eff}$, Eq.~\eqref{eq:Heff} (red circles).
For larger $\VS$ or $\gamma$, this agreement breaks down (circles not shown). 
}
\label{fig:NS}
\end{figure}

\paragraph{Scaling at confluence $U=2\Delta$.}
In Fig.~\ref{fig:NS}, we show how the current depends on $\VS/\gamma$ for several $\NS$ and $\gamma$ for the many--body confluence. 
There is a scaling regime where  
\[
\Iav \sim \frac{\VS^2}{\gamma}
\]
for all $\NS$ reachable with exact diagonalization.
This parallels the $U=0$ confluence, suggesting an extensive transport pathway that spans the whole lattice.
This is confirmed using tensor network simulations (for details see~\cite{wojtowicz_dual_2021,elenewski_performance_2021,wojtowicz_accumulative_2023}), where intensive numerical calculations allow us to reach $\NS=25$, see Fig.~\ref{fig:NS}(b). 
We fit the data, including error bars, to $\Iav= I_\infty + B \NS^{-c}$. 
When $\VS = \gamma = 0.1 \Delta$ (red crosses), the fit is very good, yielding $2 I_\infty / \gamma = 0.082 \pm 0.013$ and  $c=0.97\pm0.09$.
In words, this is a decaying (with $\NS$) diffusive component on top of a large ballistic component from an extensive bulk channel.
The existence of this ballistic component is an additional indicator of similarity between non--interacting and $U=2\Delta$ confluence points for the scaling regime. 
For $\VS=0.2\gamma$ and $\gamma = \Delta$ (blue stars), the fit yields $2 I_\infty / \gamma = 0.012 \pm 0.006$ and $c=1.1\pm0.5$. 
The fit, while still reasonable, has a larger error but is still suggestive of a ballistic plus diffusive channels. 
On the contrary, away from confluence, see Fig.~\ref{fig:NS}(a) inset, the current rapidly decays with $\NS$ as the spatially fracturing brings in multiple interfaces that interrupt current flow.
In other words, the ``flare'' contracts towards the confluence point as $\NS$ increases.

\paragraph{Conclusions.} 
We have analyzed fermionic transport through a tilted lattice with harmonically driven couplings. 
For weak interactions, there are several fractured resonances. 
These merge at a non--interacting confluence where the driving frequency matches the lattice tilt. 
Quite unexpectedly, a second confluence occurs for strong interactions, one that is not captured by the rotating wave approximation. 
This confluence occurs because the counter--rotating terms open an extensive transport channel. 
The effect is robust for the regimes studied and dominates transport for longer systems. 
These results demonstrate a novel phenomenon at the intersection of three elements---modulation, interaction, and localization---of modern quantum experiments.

\begin{acknowledgments}
We gratefully acknowledge Polish high-performance computing infrastructure PLGrid (HPC Centers: ACK Cyfronet AGH) for providing computer facilities and support within computational grant no.~PLG/2023/016370. 
This research has been supported by the National Science Centre (Poland) under project 2019/35/B/ST2/00034 (B.D.), 2020/38/E/ST3/00150 (G.W. and M.M.R.) and under the OPUS call within the WEAVE program 2021/43/I/ST3/01142 (J.Z.). G.W. acknowledges the support of the Fulbright Program. 
The research has been supported by a grant from the Priority Research Area (DigiWorld) under the Strategic Programme Excellence Initiative at Jagiellonian University (J.Z., M.M.R.). No part of this work was written by  artificial intelligence.
\end{acknowledgments}

%\bibliography{bibliography}

\begin{thebibliography}{53}%
\makeatletter
\providecommand \@ifxundefined [1]{%
 \@ifx{#1\undefined}
}%
\providecommand \@ifnum [1]{%
 \ifnum #1\expandafter \@firstoftwo
 \else \expandafter \@secondoftwo
 \fi
}%
\providecommand \@ifx [1]{%
 \ifx #1\expandafter \@firstoftwo
 \else \expandafter \@secondoftwo
 \fi
}%
\providecommand \natexlab [1]{#1}%
\providecommand \enquote  [1]{``#1''}%
\providecommand \bibnamefont  [1]{#1}%
\providecommand \bibfnamefont [1]{#1}%
\providecommand \citenamefont [1]{#1}%
\providecommand \href@noop [0]{\@secondoftwo}%
\providecommand \href [0]{\begingroup \@sanitize@url \@href}%
\providecommand \@href[1]{\@@startlink{#1}\@@href}%
\providecommand \@@href[1]{\endgroup#1\@@endlink}%
\providecommand \@sanitize@url [0]{\catcode `\\12\catcode `\$12\catcode
  `\&12\catcode `\#12\catcode `\^12\catcode `\_12\catcode `\%12\relax}%
\providecommand \@@startlink[1]{}%
\providecommand \@@endlink[0]{}%
\providecommand \url  [0]{\begingroup\@sanitize@url \@url }%
\providecommand \@url [1]{\endgroup\@href {#1}{\urlprefix }}%
\providecommand \urlprefix  [0]{URL }%
\providecommand \Eprint [0]{\href }%
\providecommand \doibase [0]{https://doi.org/}%
\providecommand \selectlanguage [0]{\@gobble}%
\providecommand \bibinfo  [0]{\@secondoftwo}%
\providecommand \bibfield  [0]{\@secondoftwo}%
\providecommand \translation [1]{[#1]}%
\providecommand \BibitemOpen [0]{}%
\providecommand \bibitemStop [0]{}%
\providecommand \bibitemNoStop [0]{.\EOS\space}%
\providecommand \EOS [0]{\spacefactor3000\relax}%
\providecommand \BibitemShut  [1]{\csname bibitem#1\endcsname}%
\let\auto@bib@innerbib\@empty
%</preamble>
\bibitem [{\citenamefont {van Nieuwenburg}\ \emph {et~al.}(2019)\citenamefont
  {van Nieuwenburg}, \citenamefont {Baum},\ and\ \citenamefont
  {Refael}}]{van_nieuwenburg_bloch_2019}%
  \BibitemOpen
  \bibfield  {author} {\bibinfo {author} {\bibfnamefont {E.}~\bibnamefont {van
  Nieuwenburg}}, \bibinfo {author} {\bibfnamefont {Y.}~\bibnamefont {Baum}},\
  and\ \bibinfo {author} {\bibfnamefont {G.}~\bibnamefont {Refael}},\
  }\bibfield  {title} {\bibinfo {title} {From {Bloch} oscillations to many-body
  localization in clean interacting systems},\ }\href
  {https://pnas.org/doi/full/10.1073/pnas.1819316116} {\bibfield  {journal}
  {\bibinfo  {journal} {Proc. Natl. Acad. Sci. U.S.A.}\ }\textbf {\bibinfo
  {volume} {116}},\ \bibinfo {pages} {9269} (\bibinfo {year}
  {2019})}\BibitemShut {NoStop}%
\bibitem [{\citenamefont {Schulz}\ \emph {et~al.}(2019)\citenamefont {Schulz},
  \citenamefont {Hooley}, \citenamefont {Moessner},\ and\ \citenamefont
  {Pollmann}}]{schulz_stark_2019}%
  \BibitemOpen
  \bibfield  {author} {\bibinfo {author} {\bibfnamefont {M.}~\bibnamefont
  {Schulz}}, \bibinfo {author} {\bibfnamefont {C.}~\bibnamefont {Hooley}},
  \bibinfo {author} {\bibfnamefont {R.}~\bibnamefont {Moessner}},\ and\
  \bibinfo {author} {\bibfnamefont {F.}~\bibnamefont {Pollmann}},\ }\bibfield
  {title} {\bibinfo {title} {Stark {Many}-{Body} localization},\ }\href
  {https://link.aps.org/doi/10.1103/PhysRevLett.122.040606} {\bibfield
  {journal} {\bibinfo  {journal} {Phys. Rev. Lett.}\ }\textbf {\bibinfo
  {volume} {122}},\ \bibinfo {pages} {040606} (\bibinfo {year}
  {2019})}\BibitemShut {NoStop}%
\bibitem [{\citenamefont {Khemani}\ \emph {et~al.}(2020)\citenamefont
  {Khemani}, \citenamefont {Hermele},\ and\ \citenamefont
  {Nandkishore}}]{khemani_localization_2020}%
  \BibitemOpen
  \bibfield  {author} {\bibinfo {author} {\bibfnamefont {V.}~\bibnamefont
  {Khemani}}, \bibinfo {author} {\bibfnamefont {M.}~\bibnamefont {Hermele}},\
  and\ \bibinfo {author} {\bibfnamefont {R.}~\bibnamefont {Nandkishore}},\
  }\bibfield  {title} {\bibinfo {title} {Localization from {Hilbert} space
  shattering: {From} theory to physical realizations},\ }\href
  {https://link.aps.org/doi/10.1103/PhysRevB.101.174204} {\bibfield  {journal}
  {\bibinfo  {journal} {Phys. Rev. B}\ }\textbf {\bibinfo {volume} {101}},\
  \bibinfo {pages} {174204} (\bibinfo {year} {2020})}\BibitemShut {NoStop}%
\bibitem [{\citenamefont {Sala}\ \emph {et~al.}(2020)\citenamefont {Sala},
  \citenamefont {Rakovszky}, \citenamefont {Verresen}, \citenamefont {Knap},\
  and\ \citenamefont {Pollmann}}]{sala_ergodicity_2020}%
  \BibitemOpen
  \bibfield  {author} {\bibinfo {author} {\bibfnamefont {P.}~\bibnamefont
  {Sala}}, \bibinfo {author} {\bibfnamefont {T.}~\bibnamefont {Rakovszky}},
  \bibinfo {author} {\bibfnamefont {R.}~\bibnamefont {Verresen}}, \bibinfo
  {author} {\bibfnamefont {M.}~\bibnamefont {Knap}},\ and\ \bibinfo {author}
  {\bibfnamefont {F.}~\bibnamefont {Pollmann}},\ }\bibfield  {title} {\bibinfo
  {title} {Ergodicity breaking arising from {Hilbert} space fragmentation in
  dipole-conserving {Hamiltonians}},\ }\href
  {https://link.aps.org/doi/10.1103/PhysRevX.10.011047} {\bibfield  {journal}
  {\bibinfo  {journal} {Phys. Rev. X}\ }\textbf {\bibinfo {volume} {10}},\
  \bibinfo {pages} {011047} (\bibinfo {year} {2020})}\BibitemShut {NoStop}%
\bibitem [{\citenamefont {Taylor}\ \emph {et~al.}(2020)\citenamefont {Taylor},
  \citenamefont {Schulz}, \citenamefont {Pollmann},\ and\ \citenamefont
  {Moessner}}]{taylor_experimental_2020}%
  \BibitemOpen
  \bibfield  {author} {\bibinfo {author} {\bibfnamefont {S.~R.}\ \bibnamefont
  {Taylor}}, \bibinfo {author} {\bibfnamefont {M.}~\bibnamefont {Schulz}},
  \bibinfo {author} {\bibfnamefont {F.}~\bibnamefont {Pollmann}},\ and\
  \bibinfo {author} {\bibfnamefont {R.}~\bibnamefont {Moessner}},\ }\bibfield
  {title} {\bibinfo {title} {Experimental probes of {Stark} many-body
  localization},\ }\href {https://link.aps.org/doi/10.1103/PhysRevB.102.054206}
  {\bibfield  {journal} {\bibinfo  {journal} {Phys. Rev. B}\ }\textbf {\bibinfo
  {volume} {102}},\ \bibinfo {pages} {054206} (\bibinfo {year}
  {2020})}\BibitemShut {NoStop}%
\bibitem [{\citenamefont {Chanda}\ \emph {et~al.}(2020)\citenamefont {Chanda},
  \citenamefont {Yao},\ and\ \citenamefont
  {Zakrzewski}}]{chanda_coexistence_2020}%
  \BibitemOpen
  \bibfield  {author} {\bibinfo {author} {\bibfnamefont {T.}~\bibnamefont
  {Chanda}}, \bibinfo {author} {\bibfnamefont {R.}~\bibnamefont {Yao}},\ and\
  \bibinfo {author} {\bibfnamefont {J.}~\bibnamefont {Zakrzewski}},\ }\bibfield
   {title} {\bibinfo {title} {Coexistence of localized and extended phases:
  {Many}-body localization in a harmonic trap},\ }\href
  {https://link.aps.org/doi/10.1103/PhysRevResearch.2.032039} {\bibfield
  {journal} {\bibinfo  {journal} {Phys. Rev. Research}\ }\textbf {\bibinfo
  {volume} {2}},\ \bibinfo {pages} {032039} (\bibinfo {year}
  {2020})}\BibitemShut {NoStop}%
\bibitem [{\citenamefont {Yao}\ \emph {et~al.}(2021{\natexlab{a}})\citenamefont
  {Yao}, \citenamefont {Chanda},\ and\ \citenamefont
  {Zakrzewski}}]{yao_many-body_2021}%
  \BibitemOpen
  \bibfield  {author} {\bibinfo {author} {\bibfnamefont {R.}~\bibnamefont
  {Yao}}, \bibinfo {author} {\bibfnamefont {T.}~\bibnamefont {Chanda}},\ and\
  \bibinfo {author} {\bibfnamefont {J.}~\bibnamefont {Zakrzewski}},\ }\bibfield
   {title} {\bibinfo {title} {Many-body localization in tilted and harmonic
  potentials},\ }\href {https://link.aps.org/doi/10.1103/PhysRevB.104.014201}
  {\bibfield  {journal} {\bibinfo  {journal} {Phys. Rev. B}\ }\textbf {\bibinfo
  {volume} {104}},\ \bibinfo {pages} {014201} (\bibinfo {year}
  {2021}{\natexlab{a}})}\BibitemShut {NoStop}%
\bibitem [{\citenamefont {Yao}\ \emph {et~al.}(2021{\natexlab{b}})\citenamefont
  {Yao}, \citenamefont {Chanda},\ and\ \citenamefont
  {Zakrzewski}}]{yao_nonergodic_2021}%
  \BibitemOpen
  \bibfield  {author} {\bibinfo {author} {\bibfnamefont {R.}~\bibnamefont
  {Yao}}, \bibinfo {author} {\bibfnamefont {T.}~\bibnamefont {Chanda}},\ and\
  \bibinfo {author} {\bibfnamefont {J.}~\bibnamefont {Zakrzewski}},\ }\bibfield
   {title} {\bibinfo {title} {Nonergodic dynamics in disorder-free
  potentials},\ }\href
  {https://linkinghub.elsevier.com/retrieve/pii/S0003491621001469} {\bibfield
  {journal} {\bibinfo  {journal} {Ann. Phys. (N. Y.)}\ }\textbf {\bibinfo
  {volume} {435}},\ \bibinfo {pages} {168540} (\bibinfo {year}
  {2021}{\natexlab{b}})}\BibitemShut {NoStop}%
\bibitem [{\citenamefont {Doggen}\ \emph {et~al.}(2021)\citenamefont {Doggen},
  \citenamefont {Gornyi},\ and\ \citenamefont {Polyakov}}]{doggen_stark_2021}%
  \BibitemOpen
  \bibfield  {author} {\bibinfo {author} {\bibfnamefont {E.~V.~H.}\
  \bibnamefont {Doggen}}, \bibinfo {author} {\bibfnamefont {I.~V.}\
  \bibnamefont {Gornyi}},\ and\ \bibinfo {author} {\bibfnamefont {D.~G.}\
  \bibnamefont {Polyakov}},\ }\bibfield  {title} {\bibinfo {title} {Stark
  many-body localization: {Evidence} for {Hilbert}-space shattering},\ }\href
  {https://link.aps.org/doi/10.1103/PhysRevB.103.L100202} {\bibfield  {journal}
  {\bibinfo  {journal} {Phys. Rev. B}\ }\textbf {\bibinfo {volume} {103}},\
  \bibinfo {pages} {L100202} (\bibinfo {year} {2021})}\BibitemShut {NoStop}%
\bibitem [{\citenamefont {Morong}\ \emph {et~al.}(2021)\citenamefont {Morong},
  \citenamefont {Liu}, \citenamefont {Becker}, \citenamefont {Collins},
  \citenamefont {Feng}, \citenamefont {Kyprianidis}, \citenamefont {Pagano},
  \citenamefont {You}, \citenamefont {Gorshkov},\ and\ \citenamefont
  {Monroe}}]{morong_observation_2021}%
  \BibitemOpen
  \bibfield  {author} {\bibinfo {author} {\bibfnamefont {W.}~\bibnamefont
  {Morong}}, \bibinfo {author} {\bibfnamefont {F.}~\bibnamefont {Liu}},
  \bibinfo {author} {\bibfnamefont {P.}~\bibnamefont {Becker}}, \bibinfo
  {author} {\bibfnamefont {K.~S.}\ \bibnamefont {Collins}}, \bibinfo {author}
  {\bibfnamefont {L.}~\bibnamefont {Feng}}, \bibinfo {author} {\bibfnamefont
  {A.}~\bibnamefont {Kyprianidis}}, \bibinfo {author} {\bibfnamefont
  {G.}~\bibnamefont {Pagano}}, \bibinfo {author} {\bibfnamefont
  {T.}~\bibnamefont {You}}, \bibinfo {author} {\bibfnamefont {A.~V.}\
  \bibnamefont {Gorshkov}},\ and\ \bibinfo {author} {\bibfnamefont
  {C.}~\bibnamefont {Monroe}},\ }\bibfield  {title} {\bibinfo {title}
  {Observation of {Stark} many-body localization without disorder},\ }\href
  {https://www.nature.com/articles/s41586-021-03988-0} {\bibfield  {journal}
  {\bibinfo  {journal} {Nature}\ }\textbf {\bibinfo {volume} {599}},\ \bibinfo
  {pages} {393} (\bibinfo {year} {2021})}\BibitemShut {NoStop}%
\bibitem [{\citenamefont {Scherg}\ \emph {et~al.}(2021)\citenamefont {Scherg},
  \citenamefont {Kohlert}, \citenamefont {Sala}, \citenamefont {Pollmann},
  \citenamefont {Hebbe~Madhusudhana}, \citenamefont {Bloch},\ and\
  \citenamefont {Aidelsburger}}]{scherg_observing_2021}%
  \BibitemOpen
  \bibfield  {author} {\bibinfo {author} {\bibfnamefont {S.}~\bibnamefont
  {Scherg}}, \bibinfo {author} {\bibfnamefont {T.}~\bibnamefont {Kohlert}},
  \bibinfo {author} {\bibfnamefont {P.}~\bibnamefont {Sala}}, \bibinfo {author}
  {\bibfnamefont {F.}~\bibnamefont {Pollmann}}, \bibinfo {author}
  {\bibfnamefont {B.}~\bibnamefont {Hebbe~Madhusudhana}}, \bibinfo {author}
  {\bibfnamefont {I.}~\bibnamefont {Bloch}},\ and\ \bibinfo {author}
  {\bibfnamefont {M.}~\bibnamefont {Aidelsburger}},\ }\bibfield  {title}
  {\bibinfo {title} {Observing non-ergodicity due to kinetic constraints in
  tilted {Fermi}-{Hubbard} chains},\ }\href
  {https://www.nature.com/articles/s41467-021-24726-0} {\bibfield  {journal}
  {\bibinfo  {journal} {Nat. Commun.}\ }\textbf {\bibinfo {volume} {12}},\
  \bibinfo {pages} {4490} (\bibinfo {year} {2021})}\BibitemShut {NoStop}%
\bibitem [{\citenamefont {Kohlert}\ \emph {et~al.}(2023)\citenamefont
  {Kohlert}, \citenamefont {Scherg}, \citenamefont {Sala}, \citenamefont
  {Pollmann}, \citenamefont {Hebbe~Madhusudhana}, \citenamefont {Bloch},\ and\
  \citenamefont {Aidelsburger}}]{kohlert_exploring_2023}%
  \BibitemOpen
  \bibfield  {author} {\bibinfo {author} {\bibfnamefont {T.}~\bibnamefont
  {Kohlert}}, \bibinfo {author} {\bibfnamefont {S.}~\bibnamefont {Scherg}},
  \bibinfo {author} {\bibfnamefont {P.}~\bibnamefont {Sala}}, \bibinfo {author}
  {\bibfnamefont {F.}~\bibnamefont {Pollmann}}, \bibinfo {author}
  {\bibfnamefont {B.}~\bibnamefont {Hebbe~Madhusudhana}}, \bibinfo {author}
  {\bibfnamefont {I.}~\bibnamefont {Bloch}},\ and\ \bibinfo {author}
  {\bibfnamefont {M.}~\bibnamefont {Aidelsburger}},\ }\bibfield  {title}
  {\bibinfo {title} {Exploring the regime of fragmentation in strongly tilted
  {Fermi}-{Hubbard} chains},\ }\href
  {https://link.aps.org/doi/10.1103/PhysRevLett.130.010201} {\bibfield
  {journal} {\bibinfo  {journal} {Phys. Rev. Lett.}\ }\textbf {\bibinfo
  {volume} {130}},\ \bibinfo {pages} {010201} (\bibinfo {year}
  {2023})}\BibitemShut {NoStop}%
\bibitem [{\citenamefont {Brantut}\ \emph {et~al.}(2012)\citenamefont
  {Brantut}, \citenamefont {Meineke}, \citenamefont {Stadler}, \citenamefont
  {Krinner},\ and\ \citenamefont {Esslinger}}]{brantut_conduction_2012}%
  \BibitemOpen
  \bibfield  {author} {\bibinfo {author} {\bibfnamefont {J.-P.}\ \bibnamefont
  {Brantut}}, \bibinfo {author} {\bibfnamefont {J.}~\bibnamefont {Meineke}},
  \bibinfo {author} {\bibfnamefont {D.}~\bibnamefont {Stadler}}, \bibinfo
  {author} {\bibfnamefont {S.}~\bibnamefont {Krinner}},\ and\ \bibinfo {author}
  {\bibfnamefont {T.}~\bibnamefont {Esslinger}},\ }\bibfield  {title} {\bibinfo
  {title} {Conduction of ultracold fermions through a mesoscopic channel},\
  }\href {https://www.sciencemag.org/lookup/doi/10.1126/science.1223175}
  {\bibfield  {journal} {\bibinfo  {journal} {Science}\ }\textbf {\bibinfo
  {volume} {337}},\ \bibinfo {pages} {1069} (\bibinfo {year}
  {2012})}\BibitemShut {NoStop}%
\bibitem [{\citenamefont {Chien}\ \emph {et~al.}(2012)\citenamefont {Chien},
  \citenamefont {Zwolak},\ and\ \citenamefont
  {Di~Ventra}}]{chien_bosonic_2012}%
  \BibitemOpen
  \bibfield  {author} {\bibinfo {author} {\bibfnamefont {C.-C.}\ \bibnamefont
  {Chien}}, \bibinfo {author} {\bibfnamefont {M.}~\bibnamefont {Zwolak}},\ and\
  \bibinfo {author} {\bibfnamefont {M.}~\bibnamefont {Di~Ventra}},\ }\bibfield
  {title} {\bibinfo {title} {Bosonic and fermionic transport phenomena of
  ultracold atoms in one-dimensional optical lattices},\ }\href
  {https://link.aps.org/doi/10.1103/PhysRevA.85.041601} {\bibfield  {journal}
  {\bibinfo  {journal} {Phys. Rev. A}\ }\textbf {\bibinfo {volume} {85}},\
  \bibinfo {pages} {041601} (\bibinfo {year} {2012})}\BibitemShut {NoStop}%
\bibitem [{\citenamefont {Brantut}\ \emph {et~al.}(2013)\citenamefont
  {Brantut}, \citenamefont {Grenier}, \citenamefont {Meineke}, \citenamefont
  {Stadler}, \citenamefont {Krinner}, \citenamefont {Kollath}, \citenamefont
  {Esslinger},\ and\ \citenamefont {Georges}}]{brantut_thermoelectric_2013}%
  \BibitemOpen
  \bibfield  {author} {\bibinfo {author} {\bibfnamefont {J.-P.}\ \bibnamefont
  {Brantut}}, \bibinfo {author} {\bibfnamefont {C.}~\bibnamefont {Grenier}},
  \bibinfo {author} {\bibfnamefont {J.}~\bibnamefont {Meineke}}, \bibinfo
  {author} {\bibfnamefont {D.}~\bibnamefont {Stadler}}, \bibinfo {author}
  {\bibfnamefont {S.}~\bibnamefont {Krinner}}, \bibinfo {author} {\bibfnamefont
  {C.}~\bibnamefont {Kollath}}, \bibinfo {author} {\bibfnamefont
  {T.}~\bibnamefont {Esslinger}},\ and\ \bibinfo {author} {\bibfnamefont
  {A.}~\bibnamefont {Georges}},\ }\bibfield  {title} {\bibinfo {title} {A
  thermoelectric heat engine with ultracold atoms},\ }\href
  {https://www.sciencemag.org/lookup/doi/10.1126/science.1242308} {\bibfield
  {journal} {\bibinfo  {journal} {Science}\ }\textbf {\bibinfo {volume}
  {342}},\ \bibinfo {pages} {713} (\bibinfo {year} {2013})}\BibitemShut
  {NoStop}%
\bibitem [{\citenamefont {Chien}\ \emph
  {et~al.}(2013{\natexlab{a}})\citenamefont {Chien}, \citenamefont {Gruss},
  \citenamefont {Ventra},\ and\ \citenamefont
  {Zwolak}}]{chien_interaction-induced_2013}%
  \BibitemOpen
  \bibfield  {author} {\bibinfo {author} {\bibfnamefont {C.-C.}\ \bibnamefont
  {Chien}}, \bibinfo {author} {\bibfnamefont {D.}~\bibnamefont {Gruss}},
  \bibinfo {author} {\bibfnamefont {M.~D.}\ \bibnamefont {Ventra}},\ and\
  \bibinfo {author} {\bibfnamefont {M.}~\bibnamefont {Zwolak}},\ }\bibfield
  {title} {\bibinfo {title} {Interaction-induced conducting–non-conducting
  transition of ultra-cold atoms in one-dimensional optical lattices},\ }\href
  {https://iopscience.iop.org/article/10.1088/1367-2630/15/6/063026} {\bibfield
   {journal} {\bibinfo  {journal} {New J. Phys.}\ }\textbf {\bibinfo {volume}
  {15}},\ \bibinfo {pages} {063026} (\bibinfo {year}
  {2013}{\natexlab{a}})}\BibitemShut {NoStop}%
\bibitem [{\citenamefont {Kohler}\ \emph {et~al.}(2005)\citenamefont {Kohler},
  \citenamefont {Lehmann},\ and\ \citenamefont {Hänggi}}]{kohler_driven_2005}%
  \BibitemOpen
  \bibfield  {author} {\bibinfo {author} {\bibfnamefont {S.}~\bibnamefont
  {Kohler}}, \bibinfo {author} {\bibfnamefont {J.}~\bibnamefont {Lehmann}},\
  and\ \bibinfo {author} {\bibfnamefont {P.}~\bibnamefont {Hänggi}},\
  }\bibfield  {title} {\bibinfo {title} {Driven quantum transport on the
  nanoscale},\ }\href
  {https://www.sciencedirect.com/science/article/pii/S0370157304005071}
  {\bibfield  {journal} {\bibinfo  {journal} {Phys. Rep.}\ }\textbf {\bibinfo
  {volume} {406}},\ \bibinfo {pages} {379} (\bibinfo {year}
  {2005})}\BibitemShut {NoStop}%
\bibitem [{\citenamefont {Krinner}\ \emph {et~al.}(2015)\citenamefont
  {Krinner}, \citenamefont {Stadler}, \citenamefont {Husmann}, \citenamefont
  {Brantut},\ and\ \citenamefont {Esslinger}}]{krinner_observation_2015}%
  \BibitemOpen
  \bibfield  {author} {\bibinfo {author} {\bibfnamefont {S.}~\bibnamefont
  {Krinner}}, \bibinfo {author} {\bibfnamefont {D.}~\bibnamefont {Stadler}},
  \bibinfo {author} {\bibfnamefont {D.}~\bibnamefont {Husmann}}, \bibinfo
  {author} {\bibfnamefont {J.-P.}\ \bibnamefont {Brantut}},\ and\ \bibinfo
  {author} {\bibfnamefont {T.}~\bibnamefont {Esslinger}},\ }\bibfield  {title}
  {\bibinfo {title} {Observation of quantized conductance in neutral matter},\
  }\href {http://www.nature.com/articles/nature14049} {\bibfield  {journal}
  {\bibinfo  {journal} {Nature}\ }\textbf {\bibinfo {volume} {517}},\ \bibinfo
  {pages} {64} (\bibinfo {year} {2015})}\BibitemShut {NoStop}%
\bibitem [{\citenamefont {Krinner}\ \emph {et~al.}(2016)\citenamefont
  {Krinner}, \citenamefont {Lebrat}, \citenamefont {Husmann}, \citenamefont
  {Grenier}, \citenamefont {Brantut},\ and\ \citenamefont
  {Esslinger}}]{krinner_mapping_2016}%
  \BibitemOpen
  \bibfield  {author} {\bibinfo {author} {\bibfnamefont {S.}~\bibnamefont
  {Krinner}}, \bibinfo {author} {\bibfnamefont {M.}~\bibnamefont {Lebrat}},
  \bibinfo {author} {\bibfnamefont {D.}~\bibnamefont {Husmann}}, \bibinfo
  {author} {\bibfnamefont {C.}~\bibnamefont {Grenier}}, \bibinfo {author}
  {\bibfnamefont {J.-P.}\ \bibnamefont {Brantut}},\ and\ \bibinfo {author}
  {\bibfnamefont {T.}~\bibnamefont {Esslinger}},\ }\bibfield  {title} {\bibinfo
  {title} {Mapping out spin and particle conductances in a quantum point
  contact},\ }\href {http://www.pnas.org/lookup/doi/10.1073/pnas.1601812113}
  {\bibfield  {journal} {\bibinfo  {journal} {Proc. Natl. Acad. Sci. U.S.A.}\
  }\textbf {\bibinfo {volume} {113}},\ \bibinfo {pages} {8144} (\bibinfo {year}
  {2016})}\BibitemShut {NoStop}%
\bibitem [{\citenamefont {Krinner}\ \emph {et~al.}(2017)\citenamefont
  {Krinner}, \citenamefont {Esslinger},\ and\ \citenamefont
  {Brantut}}]{krinner_two-terminal_2017}%
  \BibitemOpen
  \bibfield  {author} {\bibinfo {author} {\bibfnamefont {S.}~\bibnamefont
  {Krinner}}, \bibinfo {author} {\bibfnamefont {T.}~\bibnamefont {Esslinger}},\
  and\ \bibinfo {author} {\bibfnamefont {J.-P.}\ \bibnamefont {Brantut}},\
  }\bibfield  {title} {\bibinfo {title} {Two-terminal transport measurements
  with cold atoms},\ }\href
  {https://iopscience.iop.org/article/10.1088/1361-648X/aa74a1} {\bibfield
  {journal} {\bibinfo  {journal} {J. Phys.: Condens. Matter}\ }\textbf
  {\bibinfo {volume} {29}},\ \bibinfo {pages} {343003} (\bibinfo {year}
  {2017})}\BibitemShut {NoStop}%
\bibitem [{\citenamefont {Häusler}\ \emph {et~al.}(2017)\citenamefont
  {Häusler}, \citenamefont {Nakajima}, \citenamefont {Lebrat}, \citenamefont
  {Husmann}, \citenamefont {Krinner}, \citenamefont {Esslinger},\ and\
  \citenamefont {Brantut}}]{hausler_scanning_2017}%
  \BibitemOpen
  \bibfield  {author} {\bibinfo {author} {\bibfnamefont {S.}~\bibnamefont
  {Häusler}}, \bibinfo {author} {\bibfnamefont {S.}~\bibnamefont {Nakajima}},
  \bibinfo {author} {\bibfnamefont {M.}~\bibnamefont {Lebrat}}, \bibinfo
  {author} {\bibfnamefont {D.}~\bibnamefont {Husmann}}, \bibinfo {author}
  {\bibfnamefont {S.}~\bibnamefont {Krinner}}, \bibinfo {author} {\bibfnamefont
  {T.}~\bibnamefont {Esslinger}},\ and\ \bibinfo {author} {\bibfnamefont
  {J.-P.}\ \bibnamefont {Brantut}},\ }\bibfield  {title} {\bibinfo {title}
  {Scanning gate microscope for cold atomic gases},\ }\href
  {http://link.aps.org/doi/10.1103/PhysRevLett.119.030403} {\bibfield
  {journal} {\bibinfo  {journal} {Phys. Rev. Lett.}\ }\textbf {\bibinfo
  {volume} {119}},\ \bibinfo {pages} {030403} (\bibinfo {year}
  {2017})}\BibitemShut {NoStop}%
\bibitem [{\citenamefont {Gruss}\ \emph {et~al.}(2018)\citenamefont {Gruss},
  \citenamefont {Chien}, \citenamefont {Barreiro}, \citenamefont {Ventra},\
  and\ \citenamefont {Zwolak}}]{gruss_energy-resolved_2018}%
  \BibitemOpen
  \bibfield  {author} {\bibinfo {author} {\bibfnamefont {D.}~\bibnamefont
  {Gruss}}, \bibinfo {author} {\bibfnamefont {C.-C.}\ \bibnamefont {Chien}},
  \bibinfo {author} {\bibfnamefont {J.~T.}\ \bibnamefont {Barreiro}}, \bibinfo
  {author} {\bibfnamefont {M.~D.}\ \bibnamefont {Ventra}},\ and\ \bibinfo
  {author} {\bibfnamefont {M.}~\bibnamefont {Zwolak}},\ }\bibfield  {title}
  {\bibinfo {title} {An energy-resolved atomic scanning probe},\ }\href
  {http://stacks.iop.org/1367-2630/20/i=11/a=115005} {\bibfield  {journal}
  {\bibinfo  {journal} {New J. Phys.}\ }\textbf {\bibinfo {volume} {20}},\
  \bibinfo {pages} {115005} (\bibinfo {year} {2018})}\BibitemShut {NoStop}%
\bibitem [{\citenamefont {Karevski}\ and\ \citenamefont
  {Platini}(2009)}]{karevski_quantum_2009}%
  \BibitemOpen
  \bibfield  {author} {\bibinfo {author} {\bibfnamefont {D.}~\bibnamefont
  {Karevski}}\ and\ \bibinfo {author} {\bibfnamefont {T.}~\bibnamefont
  {Platini}},\ }\bibfield  {title} {\bibinfo {title} {Quantum {Nonequilibrium}
  {Steady} {States} {Induced} by {Repeated} {Interactions}},\ }\href
  {https://link.aps.org/doi/10.1103/PhysRevLett.102.207207} {\bibfield
  {journal} {\bibinfo  {journal} {Phys. Rev. Lett.}\ }\textbf {\bibinfo
  {volume} {102}},\ \bibinfo {pages} {207207} (\bibinfo {year}
  {2009})}\BibitemShut {NoStop}%
\bibitem [{\citenamefont {Prosen}(2011{\natexlab{a}})}]{prosen_exact_2011}%
  \BibitemOpen
  \bibfield  {author} {\bibinfo {author} {\bibfnamefont {T.}~\bibnamefont
  {Prosen}},\ }\bibfield  {title} {\bibinfo {title} {Exact nonequilibrium
  steady state of a strongly driven open {XXZ} chain},\ }\href
  {https://link.aps.org/doi/10.1103/PhysRevLett.107.137201} {\bibfield
  {journal} {\bibinfo  {journal} {Phys. Rev. Lett.}\ }\textbf {\bibinfo
  {volume} {107}},\ \bibinfo {pages} {137201} (\bibinfo {year}
  {2011}{\natexlab{a}})}\BibitemShut {NoStop}%
\bibitem [{\citenamefont {Prosen}(2011{\natexlab{b}})}]{prosen_open_2011}%
  \BibitemOpen
  \bibfield  {author} {\bibinfo {author} {\bibfnamefont {T.}~\bibnamefont
  {Prosen}},\ }\bibfield  {title} {\bibinfo {title} {Open {XXZ} spin chain:
  {Nonequilibrium} steady state and a strict bound on ballistic transport},\
  }\href {https://link.aps.org/doi/10.1103/PhysRevLett.106.217206} {\bibfield
  {journal} {\bibinfo  {journal} {Phys. Rev. Lett.}\ }\textbf {\bibinfo
  {volume} {106}},\ \bibinfo {pages} {217206} (\bibinfo {year}
  {2011}{\natexlab{b}})}\BibitemShut {NoStop}%
\bibitem [{\citenamefont {Žnidarič}(2011)}]{znidaric_spin_2011}%
  \BibitemOpen
  \bibfield  {author} {\bibinfo {author} {\bibfnamefont {M.}~\bibnamefont
  {Žnidarič}},\ }\bibfield  {title} {\bibinfo {title} {Spin transport in a
  one-dimensional anisotropic {Heisenberg} model},\ }\href
  {https://link.aps.org/doi/10.1103/PhysRevLett.106.220601} {\bibfield
  {journal} {\bibinfo  {journal} {Phys. Rev. Lett.}\ }\textbf {\bibinfo
  {volume} {106}},\ \bibinfo {pages} {220601} (\bibinfo {year}
  {2011})}\BibitemShut {NoStop}%
\bibitem [{\citenamefont {Karevski}\ \emph {et~al.}(2013)\citenamefont
  {Karevski}, \citenamefont {Popkov},\ and\ \citenamefont
  {Schütz}}]{karevski_exact_2013}%
  \BibitemOpen
  \bibfield  {author} {\bibinfo {author} {\bibfnamefont {D.}~\bibnamefont
  {Karevski}}, \bibinfo {author} {\bibfnamefont {V.}~\bibnamefont {Popkov}},\
  and\ \bibinfo {author} {\bibfnamefont {G.~M.}\ \bibnamefont {Schütz}},\
  }\bibfield  {title} {\bibinfo {title} {Exact matrix product solution for the
  boundary-driven {Lindblad} {XXZ} chain},\ }\href
  {https://link.aps.org/doi/10.1103/PhysRevLett.110.047201} {\bibfield
  {journal} {\bibinfo  {journal} {Phys. Rev. Lett.}\ }\textbf {\bibinfo
  {volume} {110}},\ \bibinfo {pages} {047201} (\bibinfo {year}
  {2013})}\BibitemShut {NoStop}%
\bibitem [{\citenamefont {Bertini}\ \emph {et~al.}(2021)\citenamefont
  {Bertini}, \citenamefont {Heidrich-Meisner}, \citenamefont {Karrasch},
  \citenamefont {Prosen}, \citenamefont {Steinigeweg},\ and\ \citenamefont
  {Žnidarič}}]{bertini_finite-temperature_2021}%
  \BibitemOpen
  \bibfield  {author} {\bibinfo {author} {\bibfnamefont {B.}~\bibnamefont
  {Bertini}}, \bibinfo {author} {\bibfnamefont {F.}~\bibnamefont
  {Heidrich-Meisner}}, \bibinfo {author} {\bibfnamefont {C.}~\bibnamefont
  {Karrasch}}, \bibinfo {author} {\bibfnamefont {T.}~\bibnamefont {Prosen}},
  \bibinfo {author} {\bibfnamefont {R.}~\bibnamefont {Steinigeweg}},\ and\
  \bibinfo {author} {\bibfnamefont {M.}~\bibnamefont {Žnidarič}},\ }\bibfield
   {title} {\bibinfo {title} {Finite-temperature transport in one-dimensional
  quantum lattice models},\ }\href
  {https://link.aps.org/doi/10.1103/RevModPhys.93.025003} {\bibfield  {journal}
  {\bibinfo  {journal} {Rev. Mod. Phys.}\ }\textbf {\bibinfo {volume} {93}},\
  \bibinfo {pages} {025003} (\bibinfo {year} {2021})}\BibitemShut {NoStop}%
\bibitem [{\citenamefont {Landi}\ \emph {et~al.}(2022)\citenamefont {Landi},
  \citenamefont {Poletti},\ and\ \citenamefont
  {Schaller}}]{landi_nonequilibrium_2022}%
  \BibitemOpen
  \bibfield  {author} {\bibinfo {author} {\bibfnamefont {G.~T.}\ \bibnamefont
  {Landi}}, \bibinfo {author} {\bibfnamefont {D.}~\bibnamefont {Poletti}},\
  and\ \bibinfo {author} {\bibfnamefont {G.}~\bibnamefont {Schaller}},\
  }\bibfield  {title} {\bibinfo {title} {Nonequilibrium boundary-driven quantum
  systems: {Models}, methods, and properties},\ }\href
  {https://link.aps.org/doi/10.1103/RevModPhys.94.045006} {\bibfield  {journal}
  {\bibinfo  {journal} {Rev. Mod. Phys.}\ }\textbf {\bibinfo {volume} {94}},\
  \bibinfo {pages} {045006} (\bibinfo {year} {2022})}\BibitemShut {NoStop}%
\bibitem [{\citenamefont {Kiczynski}\ \emph {et~al.}(2022)\citenamefont
  {Kiczynski}, \citenamefont {Gorman}, \citenamefont {Geng}, \citenamefont
  {Donnelly}, \citenamefont {Chung}, \citenamefont {He}, \citenamefont
  {Keizer},\ and\ \citenamefont {Simmons}}]{kiczynski_engineering_2022}%
  \BibitemOpen
  \bibfield  {author} {\bibinfo {author} {\bibfnamefont {M.}~\bibnamefont
  {Kiczynski}}, \bibinfo {author} {\bibfnamefont {S.~K.}\ \bibnamefont
  {Gorman}}, \bibinfo {author} {\bibfnamefont {H.}~\bibnamefont {Geng}},
  \bibinfo {author} {\bibfnamefont {M.~B.}\ \bibnamefont {Donnelly}}, \bibinfo
  {author} {\bibfnamefont {Y.}~\bibnamefont {Chung}}, \bibinfo {author}
  {\bibfnamefont {Y.}~\bibnamefont {He}}, \bibinfo {author} {\bibfnamefont
  {J.~G.}\ \bibnamefont {Keizer}},\ and\ \bibinfo {author} {\bibfnamefont
  {M.~Y.}\ \bibnamefont {Simmons}},\ }\bibfield  {title} {\bibinfo {title}
  {Engineering topological states in atom-based semiconductor quantum dots},\
  }\href {https://doi.org/10.1038/s41586-022-04706-0} {\bibfield  {journal}
  {\bibinfo  {journal} {Nature}\ }\textbf {\bibinfo {volume} {606}},\ \bibinfo
  {pages} {694} (\bibinfo {year} {2022})}\BibitemShut {NoStop}%
\bibitem [{\citenamefont {Wang}\ \emph {et~al.}(2022)\citenamefont {Wang},
  \citenamefont {Khatami}, \citenamefont {Fei}, \citenamefont {Wyrick},
  \citenamefont {Namboodiri}, \citenamefont {Kashid}, \citenamefont {Rigosi},
  \citenamefont {Bryant},\ and\ \citenamefont
  {Silver}}]{wang_experimental_2022}%
  \BibitemOpen
  \bibfield  {author} {\bibinfo {author} {\bibfnamefont {X.}~\bibnamefont
  {Wang}}, \bibinfo {author} {\bibfnamefont {E.}~\bibnamefont {Khatami}},
  \bibinfo {author} {\bibfnamefont {F.}~\bibnamefont {Fei}}, \bibinfo {author}
  {\bibfnamefont {J.}~\bibnamefont {Wyrick}}, \bibinfo {author} {\bibfnamefont
  {P.}~\bibnamefont {Namboodiri}}, \bibinfo {author} {\bibfnamefont
  {R.}~\bibnamefont {Kashid}}, \bibinfo {author} {\bibfnamefont {A.~F.}\
  \bibnamefont {Rigosi}}, \bibinfo {author} {\bibfnamefont {G.}~\bibnamefont
  {Bryant}},\ and\ \bibinfo {author} {\bibfnamefont {R.}~\bibnamefont
  {Silver}},\ }\bibfield  {title} {\bibinfo {title} {Experimental realization
  of an extended {Fermi}-{Hubbard} model using a {2D} lattice of dopant-based
  quantum dots},\ }\href {https://doi.org/10.1038/s41467-022-34220-w}
  {\bibfield  {journal} {\bibinfo  {journal} {Nature Communications}\ }\textbf
  {\bibinfo {volume} {13}},\ \bibinfo {pages} {6824} (\bibinfo {year}
  {2022})}\BibitemShut {NoStop}%
\bibitem [{\citenamefont {Pinho}\ \emph {et~al.}(2023)\citenamefont {Pinho},
  \citenamefont {Pires}, \citenamefont {João}, \citenamefont {Amorim},\ and\
  \citenamefont {Lopes}}]{pinho_bloch_2023}%
  \BibitemOpen
  \bibfield  {author} {\bibinfo {author} {\bibfnamefont {J.~M.~A.}\
  \bibnamefont {Pinho}}, \bibinfo {author} {\bibfnamefont {J.~P.~S.}\
  \bibnamefont {Pires}}, \bibinfo {author} {\bibfnamefont {S.~M.}\ \bibnamefont
  {João}}, \bibinfo {author} {\bibfnamefont {B.}~\bibnamefont {Amorim}},\ and\
  \bibinfo {author} {\bibfnamefont {J.~M. V.~P.}\ \bibnamefont {Lopes}},\
  }\bibfield  {title} {\bibinfo {title} {From {Bloch} oscillations to a
  steady-state current in strongly biased mesoscopic devices},\ }\href
  {https://link.aps.org/doi/10.1103/PhysRevB.108.075402} {\bibfield  {journal}
  {\bibinfo  {journal} {Phys. Rev. B}\ }\textbf {\bibinfo {volume} {108}},\
  \bibinfo {pages} {075402} (\bibinfo {year} {2023})}\BibitemShut {NoStop}%
\bibitem [{\citenamefont {Mendoza-Arenas}\ and\ \citenamefont
  {Clark}(2022)}]{mendoza-arenas_giant_2022}%
  \BibitemOpen
  \bibfield  {author} {\bibinfo {author} {\bibfnamefont {J.~J.}\ \bibnamefont
  {Mendoza-Arenas}}\ and\ \bibinfo {author} {\bibfnamefont {S.~R.}\
  \bibnamefont {Clark}},\ }\bibfield  {title} {\bibinfo {title} {Giant
  rectification in strongly-interacting boundary-driven tilted systems},\
  }\href {http://arxiv.org/abs/2209.11718} {\bibfield  {journal} {\bibinfo
  {journal} {arXiv:2209.11718}\ } (\bibinfo {year} {2022})}\BibitemShut
  {NoStop}%
\bibitem [{\citenamefont {De}\ \emph {et~al.}(2023)\citenamefont {De},
  \citenamefont {Wójtowicz}, \citenamefont {Zakrzewski}, \citenamefont
  {Zwolak},\ and\ \citenamefont {Rams}}]{de_transport_2023}%
  \BibitemOpen
  \bibfield  {author} {\bibinfo {author} {\bibfnamefont {B.}~\bibnamefont
  {De}}, \bibinfo {author} {\bibfnamefont {G.}~\bibnamefont {Wójtowicz}},
  \bibinfo {author} {\bibfnamefont {J.}~\bibnamefont {Zakrzewski}}, \bibinfo
  {author} {\bibfnamefont {M.}~\bibnamefont {Zwolak}},\ and\ \bibinfo {author}
  {\bibfnamefont {M.~M.}\ \bibnamefont {Rams}},\ }\bibfield  {title} {\bibinfo
  {title} {Transport in a periodically driven tilted lattice via the extended
  reservoir approach: {Stability} criterion for recovering the continuum
  limit},\ }\href {https://link.aps.org/doi/10.1103/PhysRevB.107.235148}
  {\bibfield  {journal} {\bibinfo  {journal} {Phys. Rev. B}\ }\textbf {\bibinfo
  {volume} {107}},\ \bibinfo {pages} {235148} (\bibinfo {year}
  {2023})}\BibitemShut {NoStop}%
\bibitem [{\citenamefont {Shirley}(1965)}]{shirley_solution_1965}%
  \BibitemOpen
  \bibfield  {author} {\bibinfo {author} {\bibfnamefont {J.~H.}\ \bibnamefont
  {Shirley}},\ }\bibfield  {title} {\bibinfo {title} {Solution of the
  {Schrödinger} equation with a {Hamiltonian} periodic in time},\ }\href
  {https://link.aps.org/doi/10.1103/PhysRev.138.B979} {\bibfield  {journal}
  {\bibinfo  {journal} {Phys. Rev.}\ }\textbf {\bibinfo {volume} {138}},\
  \bibinfo {pages} {B979} (\bibinfo {year} {1965})}\BibitemShut {NoStop}%
\bibitem [{\citenamefont {Žnidarič}(2010)}]{znidaric_matrix_2010}%
  \BibitemOpen
  \bibfield  {author} {\bibinfo {author} {\bibfnamefont {M.}~\bibnamefont
  {Žnidarič}},\ }\bibfield  {title} {\bibinfo {title} {A matrix product
  solution for a nonequilibrium steady state of an {XX} chain},\ }\href
  {https://iopscience.iop.org/article/10.1088/1751-8113/43/41/415004}
  {\bibfield  {journal} {\bibinfo  {journal} {J. Phys. A: Math. Theor.}\
  }\textbf {\bibinfo {volume} {43}},\ \bibinfo {pages} {415004} (\bibinfo
  {year} {2010})}\BibitemShut {NoStop}%
\bibitem [{\citenamefont {Gruss}\ \emph {et~al.}(2016)\citenamefont {Gruss},
  \citenamefont {Velizhanin},\ and\ \citenamefont
  {Zwolak}}]{gruss_landauers_2016}%
  \BibitemOpen
  \bibfield  {author} {\bibinfo {author} {\bibfnamefont {D.}~\bibnamefont
  {Gruss}}, \bibinfo {author} {\bibfnamefont {K.~A.}\ \bibnamefont
  {Velizhanin}},\ and\ \bibinfo {author} {\bibfnamefont {M.}~\bibnamefont
  {Zwolak}},\ }\bibfield  {title} {\bibinfo {title} {Landauer’s formula with
  finite-time relaxation: {Kramers}’ crossover in electronic transport},\
  }\href {http://www.nature.com/srep/2016/160420/srep24514/full/srep24514.html}
  {\bibfield  {journal} {\bibinfo  {journal} {Sci. Rep.}\ }\textbf {\bibinfo
  {volume} {6}},\ \bibinfo {pages} {24514} (\bibinfo {year}
  {2016})}\BibitemShut {NoStop}%
\bibitem [{\citenamefont {Elenewski}\ \emph {et~al.}(2017)\citenamefont
  {Elenewski}, \citenamefont {Gruss},\ and\ \citenamefont
  {Zwolak}}]{elenewski_communication_2017}%
  \BibitemOpen
  \bibfield  {author} {\bibinfo {author} {\bibfnamefont {J.~E.}\ \bibnamefont
  {Elenewski}}, \bibinfo {author} {\bibfnamefont {D.}~\bibnamefont {Gruss}},\
  and\ \bibinfo {author} {\bibfnamefont {M.}~\bibnamefont {Zwolak}},\
  }\bibfield  {title} {\bibinfo {title} {Communication: {Master} equations for
  electron transport: {The} limits of the {Markovian} limit},\ }\href
  {http://aip.scitation.org/doi/10.1063/1.5000747} {\bibfield  {journal}
  {\bibinfo  {journal} {J. Chem. Phys.}\ }\textbf {\bibinfo {volume} {147}},\
  \bibinfo {pages} {151101} (\bibinfo {year} {2017})}\BibitemShut {NoStop}%
\bibitem [{\citenamefont {Gruss}\ \emph {et~al.}(2017)\citenamefont {Gruss},
  \citenamefont {Smolyanitsky},\ and\ \citenamefont
  {Zwolak}}]{gruss_communication_2017}%
  \BibitemOpen
  \bibfield  {author} {\bibinfo {author} {\bibfnamefont {D.}~\bibnamefont
  {Gruss}}, \bibinfo {author} {\bibfnamefont {A.}~\bibnamefont
  {Smolyanitsky}},\ and\ \bibinfo {author} {\bibfnamefont {M.}~\bibnamefont
  {Zwolak}},\ }\bibfield  {title} {\bibinfo {title} {Communication:
  {Relaxation}-limited electronic currents in extended reservoir simulations},\
  }\href {https://aip.scitation.org/doi/abs/10.1063/1.4997022} {\bibfield
  {journal} {\bibinfo  {journal} {J. Chem. Phys.}\ }\textbf {\bibinfo {volume}
  {147}},\ \bibinfo {pages} {141102} (\bibinfo {year} {2017})}\BibitemShut
  {NoStop}%
\bibitem [{\citenamefont {Zwolak}(2020{\natexlab{a}})}]{zwolak_analytic_2020}%
  \BibitemOpen
  \bibfield  {author} {\bibinfo {author} {\bibfnamefont {M.}~\bibnamefont
  {Zwolak}},\ }\bibfield  {title} {\bibinfo {title} {Analytic expressions for
  the steady-state current with finite extended reservoirs},\ }\href
  {http://aip.scitation.org/doi/10.1063/5.0029223} {\bibfield  {journal}
  {\bibinfo  {journal} {J. Chem. Phys.}\ }\textbf {\bibinfo {volume} {153}},\
  \bibinfo {pages} {224107} (\bibinfo {year} {2020}{\natexlab{a}})}\BibitemShut
  {NoStop}%
\bibitem [{\citenamefont {Zwolak}(2020{\natexlab{b}})}]{zwolak_comment_2020}%
  \BibitemOpen
  \bibfield  {author} {\bibinfo {author} {\bibfnamefont {M.}~\bibnamefont
  {Zwolak}},\ }\bibfield  {title} {\bibinfo {title} {Comment on "{Quantum}
  transport with electronic relaxation in electrodes: {Landauer}-type formulas
  derived from the driven {Liouville}-von {Neumann} approach" [{The} {Journal}
  of {Chemical} {Physics} 153, 044103 (2020)]},\ }\href
  {http://arxiv.org/abs/2009.04466} {\bibfield  {journal} {\bibinfo  {journal}
  {arXiv:2009.04466}\ } (\bibinfo {year} {2020}{\natexlab{b}})}\BibitemShut
  {NoStop}%
\bibitem [{\citenamefont {W\'{o}jtowicz}\ \emph {et~al.}(2021)\citenamefont
  {W\'{o}jtowicz}, \citenamefont {Elenewski}, \citenamefont {Rams},\ and\
  \citenamefont {Zwolak}}]{wojtowicz_dual_2021}%
  \BibitemOpen
  \bibfield  {author} {\bibinfo {author} {\bibfnamefont {G.}~\bibnamefont
  {W\'{o}jtowicz}}, \bibinfo {author} {\bibfnamefont {J.~E.}\ \bibnamefont
  {Elenewski}}, \bibinfo {author} {\bibfnamefont {M.~M.}\ \bibnamefont
  {Rams}},\ and\ \bibinfo {author} {\bibfnamefont {M.}~\bibnamefont {Zwolak}},\
  }\bibfield  {title} {\bibinfo {title} {Dual current anomalies and quantum
  transport within extended reservoir simulations},\ }\href
  {https://link.aps.org/doi/10.1103/PhysRevB.104.165131} {\bibfield  {journal}
  {\bibinfo  {journal} {Phys. Rev. B}\ }\textbf {\bibinfo {volume} {104}},\
  \bibinfo {pages} {165131} (\bibinfo {year} {2021})}\BibitemShut {NoStop}%
\bibitem [{\citenamefont {Elenewski}\ \emph {et~al.}(2021)\citenamefont
  {Elenewski}, \citenamefont {W\'{o}jtowicz}, \citenamefont {Rams},\ and\
  \citenamefont {Zwolak}}]{elenewski_performance_2021}%
  \BibitemOpen
  \bibfield  {author} {\bibinfo {author} {\bibfnamefont {J.~E.}\ \bibnamefont
  {Elenewski}}, \bibinfo {author} {\bibfnamefont {G.}~\bibnamefont
  {W\'{o}jtowicz}}, \bibinfo {author} {\bibfnamefont {M.~M.}\ \bibnamefont
  {Rams}},\ and\ \bibinfo {author} {\bibfnamefont {M.}~\bibnamefont {Zwolak}},\
  }\bibfield  {title} {\bibinfo {title} {Performance of reservoir
  discretizations in quantum transport simulations},\ }\href
  {https://aip.scitation.org/doi/10.1063/5.0065799} {\bibfield  {journal}
  {\bibinfo  {journal} {J. Chem. Phys.}\ }\textbf {\bibinfo {volume} {155}},\
  \bibinfo {pages} {124117} (\bibinfo {year} {2021})}\BibitemShut {NoStop}%
\bibitem [{\citenamefont {Wójtowicz}\ \emph {et~al.}(2023)\citenamefont
  {Wójtowicz}, \citenamefont {Purkayastha}, \citenamefont {Zwolak},\ and\
  \citenamefont {Rams}}]{wojtowicz_accumulative_2023}%
  \BibitemOpen
  \bibfield  {author} {\bibinfo {author} {\bibfnamefont {G.}~\bibnamefont
  {Wójtowicz}}, \bibinfo {author} {\bibfnamefont {A.}~\bibnamefont
  {Purkayastha}}, \bibinfo {author} {\bibfnamefont {M.}~\bibnamefont
  {Zwolak}},\ and\ \bibinfo {author} {\bibfnamefont {M.~M.}\ \bibnamefont
  {Rams}},\ }\bibfield  {title} {\bibinfo {title} {Accumulative reservoir
  construction: {Bridging} continuously relaxed and periodically refreshed
  extended reservoirs},\ }\href
  {https://link.aps.org/doi/10.1103/PhysRevB.107.035150} {\bibfield  {journal}
  {\bibinfo  {journal} {Phys. Rev. B}\ }\textbf {\bibinfo {volume} {107}},\
  \bibinfo {pages} {035150} (\bibinfo {year} {2023})}\BibitemShut {NoStop}%
\bibitem [{\citenamefont {Velizhanin}\ \emph {et~al.}(2011)\citenamefont
  {Velizhanin}, \citenamefont {Chien}, \citenamefont {Dubi},\ and\
  \citenamefont {Zwolak}}]{velizhanin_driving_2011}%
  \BibitemOpen
  \bibfield  {author} {\bibinfo {author} {\bibfnamefont {K.~A.}\ \bibnamefont
  {Velizhanin}}, \bibinfo {author} {\bibfnamefont {C.-C.}\ \bibnamefont
  {Chien}}, \bibinfo {author} {\bibfnamefont {Y.}~\bibnamefont {Dubi}},\ and\
  \bibinfo {author} {\bibfnamefont {M.}~\bibnamefont {Zwolak}},\ }\bibfield
  {title} {\bibinfo {title} {Driving denaturation: {Nanoscale} thermal
  transport as a probe of {DNA} melting},\ }\href
  {https://link.aps.org/doi/10.1103/PhysRevE.83.050906} {\bibfield  {journal}
  {\bibinfo  {journal} {Phys. Rev. E}\ }\textbf {\bibinfo {volume} {83}},\
  \bibinfo {pages} {050906} (\bibinfo {year} {2011})}\BibitemShut {NoStop}%
\bibitem [{\citenamefont {Chien}\ \emph
  {et~al.}(2013{\natexlab{b}})\citenamefont {Chien}, \citenamefont
  {Velizhanin}, \citenamefont {Dubi},\ and\ \citenamefont
  {Zwolak}}]{chien_tunable_2013}%
  \BibitemOpen
  \bibfield  {author} {\bibinfo {author} {\bibfnamefont {C.-C.}\ \bibnamefont
  {Chien}}, \bibinfo {author} {\bibfnamefont {K.~A.}\ \bibnamefont
  {Velizhanin}}, \bibinfo {author} {\bibfnamefont {Y.}~\bibnamefont {Dubi}},\
  and\ \bibinfo {author} {\bibfnamefont {M.}~\bibnamefont {Zwolak}},\
  }\bibfield  {title} {\bibinfo {title} {Tunable thermal switching via
  {DNA}-based nano-devices},\ }\href
  {https://iopscience.iop.org/article/10.1088/0957-4484/24/9/095704} {\bibfield
   {journal} {\bibinfo  {journal} {Nanotechnology}\ }\textbf {\bibinfo {volume}
  {24}},\ \bibinfo {pages} {095704} (\bibinfo {year}
  {2013}{\natexlab{b}})}\BibitemShut {NoStop}%
\bibitem [{\citenamefont {Chien}\ \emph {et~al.}(2017)\citenamefont {Chien},
  \citenamefont {Kouachi}, \citenamefont {Velizhanin}, \citenamefont {Dubi},\
  and\ \citenamefont {Zwolak}}]{chien_thermal_2017}%
  \BibitemOpen
  \bibfield  {author} {\bibinfo {author} {\bibfnamefont {C.-C.}\ \bibnamefont
  {Chien}}, \bibinfo {author} {\bibfnamefont {S.}~\bibnamefont {Kouachi}},
  \bibinfo {author} {\bibfnamefont {K.~A.}\ \bibnamefont {Velizhanin}},
  \bibinfo {author} {\bibfnamefont {Y.}~\bibnamefont {Dubi}},\ and\ \bibinfo
  {author} {\bibfnamefont {M.}~\bibnamefont {Zwolak}},\ }\bibfield  {title}
  {\bibinfo {title} {Thermal transport in dimerized harmonic lattices: {Exact}
  solution, crossover behavior, and extended reservoirs},\ }\href
  {http://link.aps.org/doi/10.1103/PhysRevE.95.012137} {\bibfield  {journal}
  {\bibinfo  {journal} {Phys. Rev. E}\ }\textbf {\bibinfo {volume} {95}},\
  \bibinfo {pages} {012137} (\bibinfo {year} {2017})}\BibitemShut {NoStop}%
\bibitem [{\citenamefont {Chien}\ \emph {et~al.}(2018)\citenamefont {Chien},
  \citenamefont {Velizhanin}, \citenamefont {Dubi}, \citenamefont {Ilic},\ and\
  \citenamefont {Zwolak}}]{chien_topological_2018}%
  \BibitemOpen
  \bibfield  {author} {\bibinfo {author} {\bibfnamefont {C.-C.}\ \bibnamefont
  {Chien}}, \bibinfo {author} {\bibfnamefont {K.~A.}\ \bibnamefont
  {Velizhanin}}, \bibinfo {author} {\bibfnamefont {Y.}~\bibnamefont {Dubi}},
  \bibinfo {author} {\bibfnamefont {B.~R.}\ \bibnamefont {Ilic}},\ and\
  \bibinfo {author} {\bibfnamefont {M.}~\bibnamefont {Zwolak}},\ }\bibfield
  {title} {\bibinfo {title} {Topological quantization of energy transport in
  micromechanical and nanomechanical lattices},\ }\href
  {https://link.aps.org/doi/10.1103/PhysRevB.97.125425} {\bibfield  {journal}
  {\bibinfo  {journal} {Phys. Rev. B}\ }\textbf {\bibinfo {volume} {97}},\
  \bibinfo {pages} {125425} (\bibinfo {year} {2018})}\BibitemShut {NoStop}%
\bibitem [{\citenamefont {Frisk~Kockum}\ \emph {et~al.}(2019)\citenamefont
  {Frisk~Kockum}, \citenamefont {Miranowicz}, \citenamefont {De~Liberato},
  \citenamefont {Savasta},\ and\ \citenamefont
  {Nori}}]{frisk_kockum_ultrastrong_2019}%
  \BibitemOpen
  \bibfield  {author} {\bibinfo {author} {\bibfnamefont {A.}~\bibnamefont
  {Frisk~Kockum}}, \bibinfo {author} {\bibfnamefont {A.}~\bibnamefont
  {Miranowicz}}, \bibinfo {author} {\bibfnamefont {S.}~\bibnamefont
  {De~Liberato}}, \bibinfo {author} {\bibfnamefont {S.}~\bibnamefont
  {Savasta}},\ and\ \bibinfo {author} {\bibfnamefont {F.}~\bibnamefont
  {Nori}},\ }\bibfield  {title} {\bibinfo {title} {Ultrastrong coupling between
  light and matter},\ }\href
  {https://www.nature.com/articles/s42254-018-0006-2} {\bibfield  {journal}
  {\bibinfo  {journal} {Nat. Rev. Phys.}\ }\textbf {\bibinfo {volume} {1}},\
  \bibinfo {pages} {19} (\bibinfo {year} {2019})}\BibitemShut {NoStop}%
\bibitem [{\citenamefont {Hirsch}(1994)}]{Hirsch94}%
  \BibitemOpen
  \bibfield  {author} {\bibinfo {author} {\bibfnamefont {J.}~\bibnamefont
  {Hirsch}},\ }\bibfield  {title} {\bibinfo {title} {Inapplicability of the
  hubbard model for the description of real strongly correlated electrons},\
  }\href {https://doi.org/https://doi.org/10.1016/0921-4526(94)91840-6}
  {\bibfield  {journal} {\bibinfo  {journal} {Phys. B: Condens. Matter}\
  }\textbf {\bibinfo {volume} {199}},\ \bibinfo {pages} {366} (\bibinfo {year}
  {1994})}\BibitemShut {NoStop}%
\bibitem [{\citenamefont {Dutta}\ \emph {et~al.}(2015)\citenamefont {Dutta},
  \citenamefont {Gajda}, \citenamefont {Hauke}, \citenamefont {Lewenstein},
  \citenamefont {Luehmann}, \citenamefont {Malomed}, \citenamefont
  {Sowi\'{n}ski},\ and\ \citenamefont {Zakrzewski}}]{Dutta15}%
  \BibitemOpen
  \bibfield  {author} {\bibinfo {author} {\bibfnamefont {O.}~\bibnamefont
  {Dutta}}, \bibinfo {author} {\bibfnamefont {M.}~\bibnamefont {Gajda}},
  \bibinfo {author} {\bibfnamefont {P.}~\bibnamefont {Hauke}}, \bibinfo
  {author} {\bibfnamefont {M.}~\bibnamefont {Lewenstein}}, \bibinfo {author}
  {\bibfnamefont {D.-S.}\ \bibnamefont {Luehmann}}, \bibinfo {author}
  {\bibfnamefont {B.~A.}\ \bibnamefont {Malomed}}, \bibinfo {author}
  {\bibfnamefont {T.}~\bibnamefont {Sowi\'{n}ski}},\ and\ \bibinfo {author}
  {\bibfnamefont {J.}~\bibnamefont {Zakrzewski}},\ }\bibfield  {title}
  {\bibinfo {title} {Non-standard {H}ubbard models in optical lattices: a
  review},\ }\href {http://stacks.iop.org/0034-4885/78/i=6/a=066001} {\bibfield
   {journal} {\bibinfo  {journal} {Rep. Prog. Phys.}\ }\textbf {\bibinfo
  {volume} {78}},\ \bibinfo {pages} {066001} (\bibinfo {year}
  {2015})}\BibitemShut {NoStop}%
\bibitem [{\citenamefont {Landi}\ \emph {et~al.}(2014)\citenamefont {Landi},
  \citenamefont {Novais}, \citenamefont {De~Oliveira},\ and\ \citenamefont
  {Karevski}}]{landi_flux_2014}%
  \BibitemOpen
  \bibfield  {author} {\bibinfo {author} {\bibfnamefont {G.~T.}\ \bibnamefont
  {Landi}}, \bibinfo {author} {\bibfnamefont {E.}~\bibnamefont {Novais}},
  \bibinfo {author} {\bibfnamefont {M.~J.}\ \bibnamefont {De~Oliveira}},\ and\
  \bibinfo {author} {\bibfnamefont {D.}~\bibnamefont {Karevski}},\ }\bibfield
  {title} {\bibinfo {title} {Flux rectification in the quantum {X} {X} {Z}
  chain},\ }\href {https://link.aps.org/doi/10.1103/PhysRevE.90.042142}
  {\bibfield  {journal} {\bibinfo  {journal} {Phys. Rev. E}\ }\textbf {\bibinfo
  {volume} {90}},\ \bibinfo {pages} {042142} (\bibinfo {year}
  {2014})}\BibitemShut {NoStop}%
\bibitem [{\citenamefont {Balachandran}\ \emph {et~al.}(2018)\citenamefont
  {Balachandran}, \citenamefont {Benenti}, \citenamefont {Pereira},
  \citenamefont {Casati},\ and\ \citenamefont
  {Poletti}}]{balachandran_perfect_2018}%
  \BibitemOpen
  \bibfield  {author} {\bibinfo {author} {\bibfnamefont {V.}~\bibnamefont
  {Balachandran}}, \bibinfo {author} {\bibfnamefont {G.}~\bibnamefont
  {Benenti}}, \bibinfo {author} {\bibfnamefont {E.}~\bibnamefont {Pereira}},
  \bibinfo {author} {\bibfnamefont {G.}~\bibnamefont {Casati}},\ and\ \bibinfo
  {author} {\bibfnamefont {D.}~\bibnamefont {Poletti}},\ }\bibfield  {title}
  {\bibinfo {title} {Perfect {Diode} in {Quantum} {Spin} {Chains}},\ }\href
  {https://link.aps.org/doi/10.1103/PhysRevLett.120.200603} {\bibfield
  {journal} {\bibinfo  {journal} {Phys. Rev. Lett.}\ }\textbf {\bibinfo
  {volume} {120}},\ \bibinfo {pages} {200603} (\bibinfo {year}
  {2018})}\BibitemShut {NoStop}%
\end{thebibliography}

\end{document}